\documentclass[useAMS,usegraphicx]{mn2e}

\usepackage{times}

\newcommand{\wgamma}{\ensuremath{\Gamma}}

\newcommand{\ecut}{\ensuremath{E_{\rm{cut}}}}
\newcommand{\efold}{\ensuremath{E_{\rm{fold}}}}
\newcommand{\ecy}[1]{\ensuremath{E_{\rm{c#1}}}}
\newcommand{\wcy}[1]{\ensuremath{\sigma_{\rm{c#1}}}}
\newcommand{\dcy}[1]{\ensuremath{\tau_{\rm{c#1}}}}

\newcommand{\igr}{IGR J18179-1621 }

\usepackage{graphicx}
\usepackage{url}
\usepackage{lscape}
\usepackage[T1]{fontenc}
\usepackage{txfonts}

\title[\emph{INTEGRAL} and \emph{Swift/XRT} observations on IGR J18179-1621]{\emph{INTEGRAL} and \emph{Swift/XRT} observations on IGR J18179-1621}
\author[J. Li, S. Zhang, D. F. Torres, A. Papitto, Y. P. Chen and J. M. Wang]{J. Li$^{1,2}$\footnotemark[1], S. Zhang$^{1}$\thanks{E-mail: jianli@ihep.ac.cn (JL); szhang@ihep.ac.cn (SZ)}, D. F. Torres$^{2,3}$, A. Papitto$^{2}$, Y. P. Chen$^{1}$ and J. M. Wang$^{1,4,5}$\\
$^{1}$ Key Laboratory for Particle Astrophysics, Institute of High Energy Physics, Chinese Academy of Sciences, 19B Yuquan Road, Beijing 100049, China\\
$^{2}$Institut de Ci\`encies de l'Espai (IEEC-CSIC),
              Campus UAB,  Torre C5, 2a planta,
              08193 Barcelona, Spain\\
$^{3}$Instituci\'o Catalana de Recerca i Estudis Avan\c{c}ats (ICREA).\\
$^{4}$National Astronomical Observatories of China, Chinese Academy of Sciences, 20A Datun Road, Beijing 100020, China\\
$^{5}$Theoretical Physics Center for Science Facilities (TPCSF), Chinese Academy of Sciences, Beijing 100049, China\\}
\begin{document}

\date{Accepted  ??? ; Received ??? }

\pagerange{\pageref{firstpage}--\pageref{lastpage}} \pubyear{2002}

\maketitle

\label{firstpage}

\begin{abstract}

\igr is a hard X-ray binary transient discovered recently by \emph{INTEGRAL}. Here we report on detailed timing and spectral analysis on IGR J18179-1621 in X-rays based on available \emph{INTEGRAL} and \emph{Swift} data. From the \emph{INTEGRAL} analysis, \igr is detected with a significance of 21.6~$\sigma$ in the 18--40 keV band by \emph{ISGRI} and 15.3~$\sigma$ in the 3--25 keV band by \emph{JEM-X}, between 2012-02-29 and 2012-03-01. We analyze two quasi-simultaneous \emph{Swift} ToO observations. A clear 11.82 seconds pulsation is detected above the white noise at a confidence level larger than $99.99\%$. The pulse fraction is estimated as $22\pm8\%$ in 0.2-10 keV. No sign of pulsation is detected by \emph{INTEGRAL/ISGRI} in the 18--40 keV band. With \emph{Swift} and \emph{INTEGRAL} spectra combined in soft and hard X-rays, \igr could be fitted by an absorbed power law with a high energy cutoff plus a Gaussian absorption line centered at 21.5 keV. An additional absorption intrinsic to the source is found, while the absorption line is evidence for most probably originated from cyclotron resonant scattering and suggests a magnetic field in the emitting region of $\sim$ $2.4\times10^{12}$ Gauss.

\end{abstract}


\section{Introduction}

One of the most effective techniques to estimate the strength of the
magnetic field of a neutron star is the detection of cyclotron
resonant scattering features (CRSF) in its X-ray spectrum. The fundamental electron cyclotron resonance energy is
$E=11.6\; B_{12}(1+z_g)^{-1}$ keV, where $B_{12}$ is the magnetic
field strength of the neutron star in units of $10^{12}$ G, and $z_g$ is the gravitational redshift.
The surface magnetic field of neutron stars in accreting X-ray binaries can then be determined
through the observation of the CRSF, which shows absorption lines at the fundamental electron cyclotron resonance and its
high-energy harmonics.

So far, CRSF are identified in 17 X-ray binaries, and show hints in another 10
(Pottschmidt et al. 2011,  Makishima et al 1999,  Coburn et al 2001,  Yamamoto et al 2011). All the X-ray binaries with identified or possible CRSF are high mass X-ray binaries, except for 4U 1626-67, and all host X-ray pulsars. 7 out of the 10 candidates host X-ray pulsars too, with the exceptions of  XTE J1739-302, 4U 1700-377, and IGR 16318-4848, for which pulsations are not yet detected. The magnetic field strength of CRSF-identified X-ray binaries cluster in a relatively narrow range of $(1.1-6.2)\times10^{12}$ Gauss (assuming the typical neutron star parameters, $z_g$ $\sim$0.3).

 \igr is a newly discovered hard X-ray binary transient found by \emph{INTEGRAL} during inner Galactic disk observations performed on
 2012-02-29 -- MJD 55986 (see the ATel by Tuerler et al. 2012).
%
The significant detection of \igr by \emph{INTEGRAL} reveals an absorption line $\sim$20.8 keV, which may result from cyclotron resonant scattering (Tuerler et al. 2012). In subsequent \emph{Swift/XRT} and \emph{Fermi/GBM} observations (see the ATels by Halpern et al. 2012, Li et al.2012, and Finger et al. 2012) a 11.82 s pulsation was discovered. The absorption line and the 11.82~s pulsation suggested that \igr is a high mass X-ray binary hosting a pulsar. Due to the overlap between the \emph{Swift} position of
the X-ray source and 2MASS J18175218-1621316, Li et al. (2012)
proposed the latter as the infrared counterpart of IGR J18179-1621. Such a correlation is compatible with the \emph{Chandra}
determination of the position, later obtained by Paizis et al. (2012).
Here, we report on spectra and timing analysis of the \emph{INTEGRAL} observations as well as on two quasi-simultaneous \emph{Swift} ToO observations.

\section{Observations and Data analysis}
\emph{INTEGRAL} (Winkler et al. 2003) is a $\gamma$-ray mission optimized to work between 15 keV--10 MeV. Its main instruments are the \emph{IBIS} (15 keV--10 MeV; Ubertini et al. 2003) and the Spectrometer
on board \emph{INTEGRAL} (SPI, 20 keV--8 MeV; Vedrenne et al. 2003).
The \emph{INTEGRAL} observations were
carried out in individual Science Windows (ScWs), which have a typical time
duration of about 2000 s.

     For \emph{INTEGRAL} analysis, we use \emph{IBIS/ISGRI} and \emph{JEM--X} data accumulated from the inner Galactic disk observations performed on 2012-02-29 (MJD 55986) and 2012-03-01 (MJD 55987). Our data set
comprise 50 Science Windows covering revolution 1145, adding up to a total exposure time of 88.2 ks in \emph{IBIS/ISGRI} and
39.4 ks in \emph{JEM--X} (19.9 ks in \emph{JEM--X1} and 19.5 ks in \emph{JEM--X2}).
The data reduction is performed using the standard ISDC offline scientific analysis software version 9.0. We use the
latest instrument characteristics in our analysis and follow the position of \igr determined by Paizis et al. (2012).
An \emph{IBIS/ISGRI} image for each ScW is generated in the 18--40 keV energy band.
 The spectrum and lightcurve of \igr are produced following standard steps as explained in \emph{IBIS} Analysis User Manual, running the pipeline from the raw to the SPE and LCR level. Photon events are extracted following the standard procedure as explained in the \emph{IBIS} Analysis User Manual for timing analysis. The spectrum of \igr from JEM--X observations are similarly produced following the standard steps using OSA 9.0.

\emph{Swift} (Gehrels et al. 2004) is a $\gamma$-ray burst explorer
launched on November 20, 2004. It carries three co-aligned detectors: the Burst Alert Telescope (\emph{BAT}, Barthelmy et al. 2005),
the X-Ray Telescope (\emph{XRT}, Burrows et al. 2005), and the
Ultraviolet/Optical Telescope (\emph{UVOT}, Roming et al. 2005).
The \emph{XRT} uses a grazing incidence Wolter I telescope. \emph{XRT} has an
effective area of 110 cm$^2$, a FOV of 23.6 arcmin, an angular resolution
 (half-power diameter) of 15 arcsec, and it operates in the
0.2--10 keV energy range, providing the possibility of extending
the investigation of the source to soft X-rays.

Following the discovery of \igr (Tuerler et al. 2012) on 2012-02-29, we acquired two \emph{Swift} ToO observations quasi-simultaneously with \emph{INTEGRAL}, which were executed on 2012-02-29 (MJD 55986, ID 00032293001) and
 2012-03-01 (MJD 55986, ID 00032293002) leading to an exposure of $\sim$1.9 ks and $\sim$2 ks, respectively.
These two observations were carried out by
\emph{Swift/XRT} in Photon Counting (PC) mode.
We select PC data with event grades 0-12 (Burrows et al. 2005). Due to the relatively
high count rate observed by \emph{Swift/XRT} ($\sim$1.9 counts/s), the observation is affected by pile-up. To correct for the pile-up effect we estimate the size of the Point Spread Function (PSF) core affected. By comparing the observed and nominal PSF (Romano et al. 2006; Vaughan et al. 2006), a radius of 4 pixels are determined and all the data within this radius
from \igr are excluded. Source events are accumulated within an annulus (inner radius of 4 pixels and outer radius of 30 pixels, 1 pixel $\sim$ 2.36 arcsec)\footnote{Please see \url{http://www.swift.ac.uk/pileupthread.shtml} for more information}. Background events are accumulated within a circular, source-free
region with a radius of 60 pixels.
For timing analysis, BARYCORR task is used to perform barycentric corrections to the photon arrival times using the Chandra position given by Paizis et al. (2012). We extract lightcurves with a time resolution of 2.5 seconds. XRTLCCORR task is used to account for the pile-up correction in the background-subtracted light curves.
For our spectral analysis, we extract events in the same regions as those adopted for the lightcurve creation.
Exposure maps are generated with the task XRTEXPOMAP. Ancillary response files are
generated with XRTMKARF, to account for different extraction regions,
vignetting, and PSF corrections. We analyze the \emph{Swift/XRT} 0.2--10 keV data by
HEAsoft version 6.9 and spectral fitting is performed using XSPEC V.12.6.0. The error for timing and spectral analysis are estimated at 90$\%$ confidence level.

\section{Timing and Spectroscopy}
\begin{figure}
\includegraphics[width=90mm]{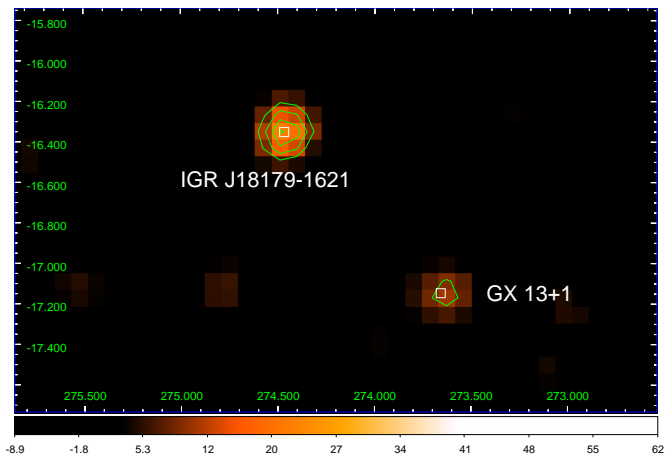}
\caption{The 18--40 keV \emph{INTEGRAL/ISGRI} significance mosaic image of \igr sky region, derived by combining all
\emph{INTEGRAL/ISGRI} data in the 18--40 keV energy band. The significance level is given
by the color scale which is linear. Corresponding significance and color could be found in the lower color bar.
The contours of significance start at 9 $\sigma$, and following steps of 4 $\sigma$. X-axis and Y-axis are RA and DEC in the unit of degree. }
\end{figure}

Swift imaging analysis shows a single source detected at soft X-rays. Using the online Swift tool\footnote{Please see \url{http://www.swift.ac.uk/user_objects/}for more information}
we find a position for \igr at  RA = 274.4675 (18h 17m 52.20s), DEC = -16.3589 (-16d 21' 31.9'') (J2000),
with a 90$\%$ confidence error of 2.2 arcsec, consistent with the position measured by \emph{INTEGRAL/JEM-X} (Tuerler et al. 2012) and \emph{Chandra} (Paizis et al. 2012). This position determination
uses a Point Spread Function fit, with bad column and pile-up correction built in. For the enhanced positions the absolute astrometry is corrected using field stars in
the \emph{UVOT}. We have used the \emph{XRT-UVOT} alignment, and matching \emph{UVOT} field sources to the USNO-B1 catalogue (Evans et al. 2009).
With all \emph{INTEGRAL/ISGRI} data combined, \igr is detected with a significance level of 21.6 $\sigma$ 
in the 18--40 keV energy band, over a 88.2 ks exposure. Figure 1 shows the \emph{INTEGRAL/ISGRI} significance mosaic image of the \igr sky region in 18-40 keV.  \igr is also consistently
 detected by \emph{JEM--X1} and \emph{JEM--X2}. Combining all \emph{JEM--X1} data, \igr is detected with a significance level of 11.4~$\sigma$ in the 3--10 keV band, 15.9~$\sigma$ in the 10--25 keV band, and 15.3~$\sigma$
in the 3-25 keV band, under a exposure of 19.9 ks. \emph{JEM--X2} detects \igr at a similar level.
The significant detection of \igr by \emph{Swift/XRT} and \emph{INTEGRAL} allows for a detailed spectral and timing analysis in both soft and hard X-rays.

\subsection{Timing analysis}

\emph{Swift/XRT} lightcurves in 2.5 seconds bin are produced in the 0.2--10 keV band for the two ToO observations mentioned above. A constant fit to the lightcurve yields an average count rate of 0.91$\pm$0.03 counts/s and a reduced $\chi^{2}$ of 1.03 for 1238 d.o.f. In order to search for any periodic signal in the \emph{Swift/XRT} lightcurve, we
use the Lomb--Scargle periodogram method (Lomb 1976; Scargle
1982). Power spectra are generated for the lightcurve
using the PERIOD subroutine (Press $\&$ Rybicki 1989). The 99.99\%
white noise significance levels are estimated
using Monte Carlo simulations (see e.g. Kong, Charles $\&$
Kuulkers 1998). A significant signal at $11.82\pm0.03$ s is detected above
the white noise at a confidence level larger than $99.99\%$ (figure 2, upper panel). The pulsation exhibits a single peak profile and pulse fraction of $22\pm8\%$ (figure 2, middle panel). The signal we detect is consistent with the results
reported by Halpern et al. (2012), Li et al. (2012), and Finger et al. (2012).

The \emph{INTEGRAL/ISGRI} lightcurve is produced by running the pipeline to LCR level and binned with 1800 seconds. Fitting a constant to the \emph{INTEGRAL/ISGRI} lightcurve yields an average count rate of
 3.10$\pm$0.16 counts/s and a reduced $\chi^{2}$ of 1.01 for 53 d.o.f. For timing analysis in hard X-ray, we extracted all \emph{INTEGRAL/ISGRI} photon events from IGR J18179-1621 in the 18--40 keV band. For a larger
 signal-to-noise ratio, only photons from fully illuminated
pixels (Pixel Illuminated Factor=1) are included. Barycentric corrections
 are applied to the photon arrival times using the \emph{Chandra} position given by Paizis et al. (2012), and the pulsation is searched for by using the \emph{Xronos} routine \emph{efsearch}.
 Fourier period resolution (FPR),
$\delta P = P^{2}/2Tobs$, is used as search resolution. With $P=11.82$ s and
$2Tobs = 2$ d, we obtain $\delta P =4\times10^{-4}$ s. To search within the interval defined
by the 3~$\sigma$ error on the \emph{Swift/XRT} determination of the period (i.e.,
$2\times3\times0.01=0.06$ s), we perform 150 period trails. The trail periods are sampled using 8 phase bins.
For a $\chi^{2}$ distribution with 7 degree of
freedom (see, e.g. Leahy et al. 1983) the detection
at a 3~$\sigma$ confidence level for a single trial is
$\chi^2=21.8$. Considering the number of trials performed, 3 $\sigma$ confidence level
increases to $\chi^2=33.9$. The results are shown in the lower panel of figure 2.
The most significant peak has a $\chi^{2}=20.6$, which therefore does not allow to
draw a detection of the signal in the \emph{INTEGRAL/ISGRI} events. This is most
probably the result of background dominated nature of \emph{INTEGRAL/ISGRI}
data. We also note how this analysis assumes that the possible orbital
period of the system does not significantly affect the frequency of
the signal, over the considered time-interval (i.e., is > 20 d).

 \begin{figure}
   \centering
\includegraphics[width=84mm]{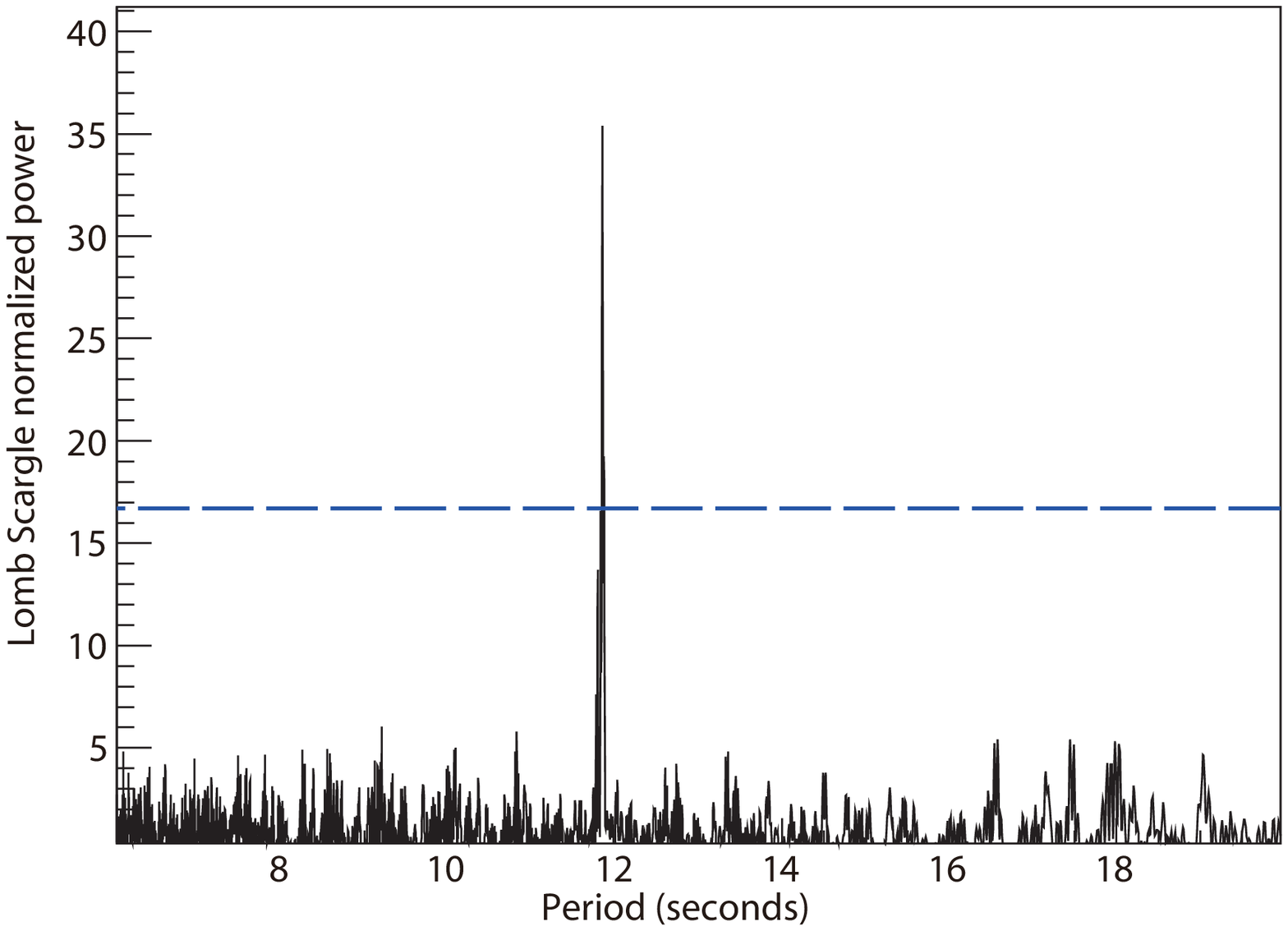}
\includegraphics[width=84mm]{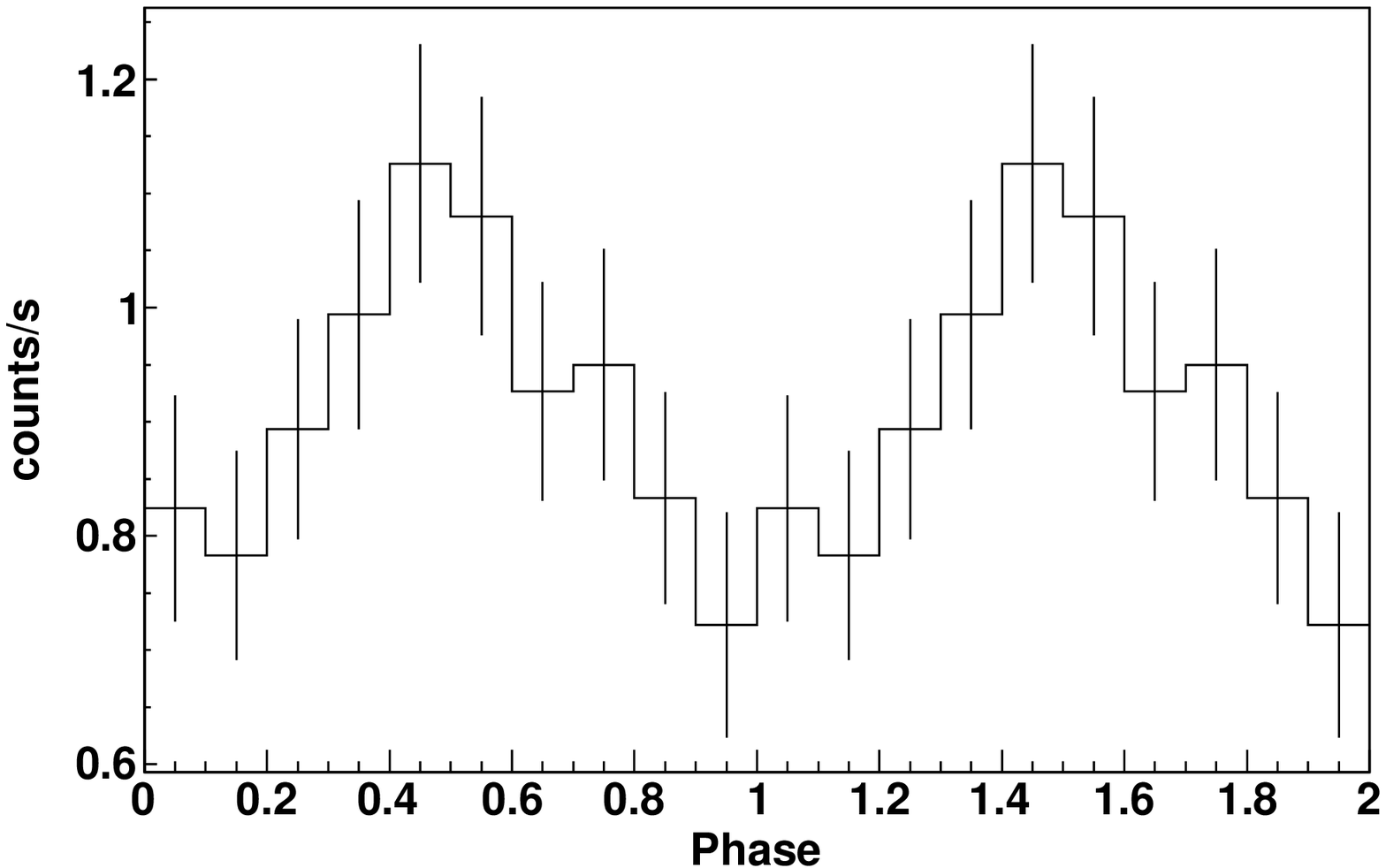}

\includegraphics[width=84mm]{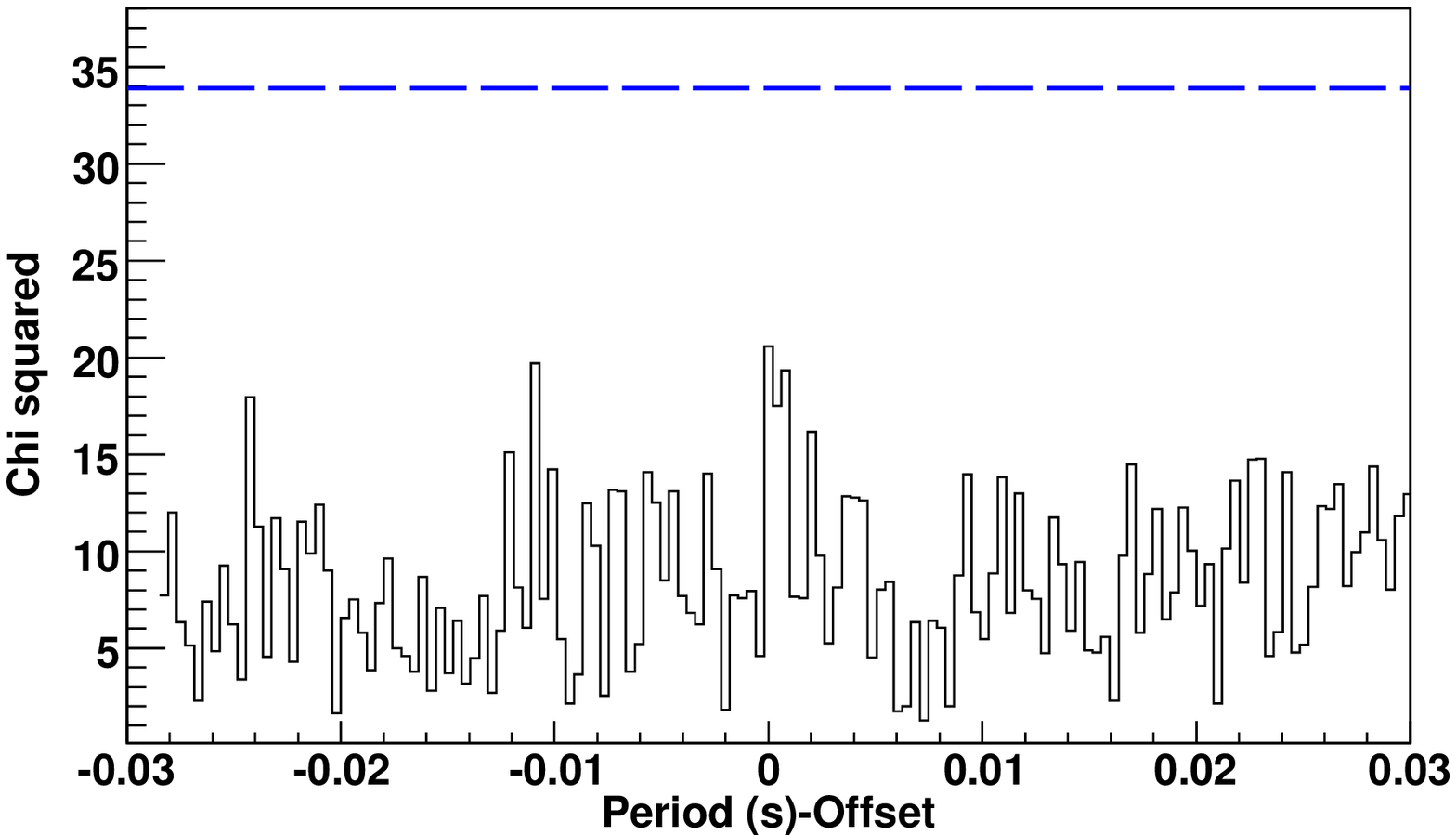}
   \caption{
Upper: Lomb-Scargle normalized power spectrum of \igr from 2.5 s \emph{Swift/XRT}-binned lightcurve in 0.2-10 keV. A peak at 11.82 seconds is detected.
The dotted blue line is the 99.99$\%$ white--noise confidence level. Middle: pulse profile of \igr in 0.2-10 keV from \emph{Swift/XRT}.
Lower: $\chi^{2}$ results from \textit{efsearch} of \emph{INTEGRAL/ISGRI} photons events. The best period is 11.81798 s and the X-axis is
the period offset to it, which ``0" corresponding to the best period. The dotted blue line is 3 $\sigma$ confidence level  }
              \label{FigGam}%
    \end{figure}

\subsection{Spectral analysis}

\emph{INTERGAL} and \emph{Swift} observations of \igr are quasi-simultaneous within continuous two days. Consequently it provides a chance for simultaneous spectral analysis in soft and hard X-rays. Combined with \emph{Swift/XRT} and \emph{INTEGRAL/JEMX $\&$ ISGRI}, we carry out systematic spectral analysis for MJD 55986 (2012-02-29) and MJD 55987 (2012-02-29) separately. We checked a
posteriori that the analysis on the two days separately gave consistent results. The spectra from the two consecutive days are then combined to increase the statistics and better constrain the parameters. To fit simultaneously the spectra
of various instruments  --\emph{Swift/XRT} (0.2--10 keV), \emph{INTEGRAL/JEMX-1} (5--30 keV), \emph{JEMX-2} (5--30 keV) and \emph{ISGRI} (18--50 keV)--  we introduce normalization constants, which are fixed to 1 for the
\emph{JEMX-1} spectrum, and free to vary for the others. Using an absorbed power-law model, we obtained a reduced $\chi^{2}$ = 4.02 over 118 degrees of freedom, which is far from being acceptable. To improve the fitting, we followed Coburn et al. 2002, adding a
high energy cutoff (PLCUT; White et al. 1983) to the absorbed powerlaw. The analytic form of the model is:

\begin{equation}
{\ensuremath{
   \mathrm{PLCUT}(E) = A\ E^{-\Gamma}\times\
   \cases{1&$(E\leq\ecut)$ \cr {\rm e}^{-(E-\ecut)/\efold}&$(E>\ecut)$}
}}
\end{equation}
where \wgamma\ is the photon index, and \ecut\ and \efold\ are the cutoff and folding energies, respectively. Applying PLCUT model an apparent structure locates at $\sim$20 keV
in the residuals (Figure 3, lower panel), which indicates the presence of a CRSF. To model the possible CRSF we use a Gaussian-shaped absorption feature (GABS), following Coburn et al. 2002. The functional form for it is:
\begin{equation}
{\ensuremath{
   {\rm GABS}(E) = \dcy{}\
   {\rm e}^{-(E-\ecy{})^{2}/(2\wcy{}^{2})}.
}}
\end{equation}
In this model \ecy{} is the energy of the resonance, \dcy{} is the depth of the line at the resonance, and \wcy{} is the width. In addition to model the continuum with PLCUT, a GABS is added to fit the CRSF. We obtain an improvement in the fit quality with small uniformly distributed residuals (Figure 3, upper panel). An CRSF is significantly discovered at $21.5_{-0.9}^{+1.0}$ keV. F-test analysis gives a probability of $ 8.1 \times 10^{-9}$ that the improvement of the fit occurs by chance.
The centroid energy of the CRSF we detect lies close to the energy at which the \emph{JEMX-1/2} and the \emph{ISGRI} spectra overlap; thus we cannot rule out how the observed shape could be
partly influenced by the cross-calibration between the two instruments.
Parameters derived from the spectral fitting are shown in Table 1.

 \begin{figure}
   \centering
\includegraphics[width=82mm,angle=0]{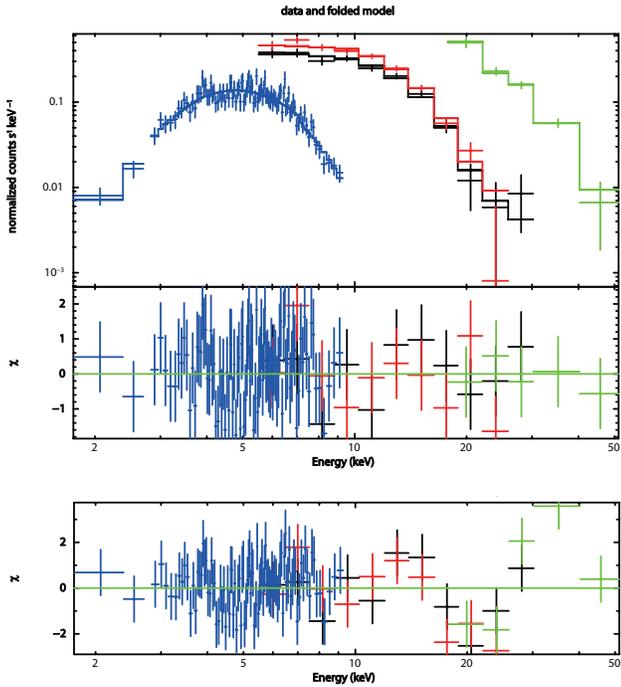}
   \caption{Upper: Combined folded spectra from \emph{Swift/XRT} (blue), \emph{INTEGRAL/JEMX--1} (black), \emph{JEMX--2} (red) and \emph{ISGRI} (green). The spectra are fitted with PLCUT plus GABS model and residuals
   between data and model are shown. Lower: residuals
   between data and model in the case of PLCUT only. A structure at $\sim$20 kev
 indicates the presence of a CRSF}
    \label{FigGam}%
    \end{figure}

\begin{table*}
\centering
\caption{Fiting Parameters of the combined spectra from \emph{Swift} and \emph{INTEGRAL}}             
\label{table1}      
\centering                          
\begin{tabular}{cccccccccc}        
\hline\hline                 
 Hydrogen  & Photon & Cutoff  & Folding     &Centroid        & Absorption Line    & Absorption Line & reduced $\chi^{2}$ (D.O.F) \\
column density &  Index  & Energy &  Energy    &Energy          &                  &    &            \\
$10^{22}$ cm$^{-2}$&   & \ecut (keV) &\efold (keV)&\ecy (keV)  &   \ensuremath{\sigma_{\rm{c}}}(keV)      &    \ensuremath{\tau_{\rm{c}}} &                      \\
\hline
                \\     
$12.3_{-1.7}^{+1.8}$  & $0.59_{-0.25}^{+0.25}$&$9.97_{-0.94}^{+0.97}$ & $6.10_{-0.58}^{+0.70}$ &$21.5_{-0.9}^{+1.0}$ &$2.61_{-0.88}^{+1.14}$ &$6.93_{-2.11}^{+2.44}$ & 0.923 (113)\\
\\
\hline
\\
Normalization factors&&&&& &&&&\\
\\
\emph{INTEGRAL}&                                                &\emph{Swift}&\\
\emph{JEM X--2}/\emph{JEM X--1}  &\emph{ISGRI }/\emph{JEM X--1} &\emph{XRT}/\emph{JEM X--1} &\\
\\
\hline
$1.21_{-0.08}^{+0.08}$&$1.17_{-0.25}^{+0.31}$&$0.67_{-0.06}^{+0.07}$&\\
\hline
\hline                                   
\end{tabular}
\end{table*}

\section{Discussion}

We present timing and spectral analysis of \igr from quasi-simultaneous observations made by \emph{Swift/XRT} and \emph{INTEGRAL}. A 11.82 s pulsation is discovered in \emph{Swift/XRT} lightcurve in the 0.2--10 keV band, which is consistent with previous reports by Halpern et al. 2012, Li et al. 2012, and Finger et al. 2012. Because of more data being included in our analysis, a higher significance detection is obtained when compared with that of Tuerler et al. 2012. The 11.82~s pulsation is searched within \emph{INTEGRAL/ISGRI} data in the 18--40 keV band, but results in a non-detection. Using quasi-simultaneous data from \emph{Swift} and \emph{INTEGRAL}, for the first time we carry out a combined spectral analysis in soft and hard X-rays. Above a continuum described by a $\sim $0.6 power-law with a cut-off at $\sim$10 keV, a CRSF is significantly detected at 21.5 keV,
which is consistent with, but more precise and constrained than the
results by Tuerler et al. 2012, obtained using \emph{INTEGRAL} data
only. Among those X-ray binaries which show CRSF and which are fitted by the same model (Coburn et al. 2002), \igr have a moderate slope. \ecut and \efold
are low comparing with other sources. 4U 0115+63 is a transient X-ray pulsar too with 3.6~s pulsation and 24-d orbit. It have a similar cutoff energy (10 keV) and folding energy (9.3 keV) than IGR J18179-1621. However, since little is known about IGR J18179-1621, we cannot draw further conclusions on the similarity between these two sources. Though IGR J18179-1621's 2.61 keV width of CRSF is not uncommon, its optical depth of 6.3 is the largest --about ten times that of other sources. A correlation between CRSF relative width (\ensuremath{\sigma_{\rm{c}}}/\ensuremath{E_{\rm{c}}}) and optical depth (\ensuremath{\tau_{\rm{c}}}) is observed (Coburn et al. 2002). However, because of the unusually large CRSF optical depth, \igr is far from the correlation. On the contrary, \igr follows another two correlations with other sources, CRSF width (\ensuremath{\sigma_{\rm{c}}}) versus centroid energy (\ensuremath{E_{\rm{c}}}), and cutoff energy (\ecut) versus centroid energy (\ensuremath{E_{\rm{c}}}) (Coburn et al. 2002).
An absorption ($12.3\times10^{22}$ cm$^{-2}$) which is much larger than the Galactic column density ($1.2\times10^{22}$ cm$^{-2}$) at IGR J18179-1621's position is obtained. This possibly indicates an additional absorption intrinsic to the source. Li et al. 2012 proposed 2MASS J18175218-1621316 as the infrared counterpart for IGR J18179--1621. This source is obscured and it is   well measured only in the Ks band: Ks magnitude=11.14. Both may indicate a complicated surrounding of IGR J18179--1621.

Given the value of the spin period, the detection of an absorption feature compatiable with a CRSF, and the proposed optical counterpart, \igr is most plausibly an accreting pulsar. In general, the transient behavior of X-ray binaries is powered by accretion of matter from the companion to the magnetic poles of the neutron star. The accretion flow onto the magnetic pole will be decelerated in a radiative shock above the neutron star surface when the luminosity
 reaches $10^{37}$ erg s$^{-1} $   (Basko $\&$ Sunyaev 1976). Radiation will be modulated as a fan beam coming out from the bottom region of the shock and peaked
perpendicular to the magnetic axis. In a lower accretion rate, while luminosity is less than $10^{37}$ erg s$^{-1} $, radiation will be formed into a pencil--beam,
which the maximum direction is along the magnetic axis. During a transient outburst similar with the one lead to discovery of IGR J18179--1621, both the fan--beam and pencil beam will influence the pulse profiles in the
lightcurve. In a high (low) accretion phase, the fan--beam (pencil) component is dominating the emission region, leading to a double (single) peak pulse profile.
The transition point between these two different phases is $\sim 10^{37}$ erg s$^{-1} $. A demonstration of the pulse profile transformation from two peaks to a single
 peak accompanying the luminosity evolution is seen in V0332+53 (see, e.g., Zhang et al. 2005). \igr is characterized with a single peak pulse profile and its unabsorbed flux in 1.5--50
keV is $\sim 1.3\times10^{-9}$ erg cm$^{-2}$ s$^{-1}$. If we apply L $<$ $10^{37}$ erg s$^{-1} $ as the threshold of pulse profile shift, we obtain a upper limit on
the \igr distance at d $<$ 8 kpc.

From the combined \emph{Swift} and \emph{INTEGRAL} spectra CRSF of \igr is identified at 21.5 keV and no high energy harmonics are discovered. If the fundamental electron cyclotron resonance energy is 21.5 keV, this will indicate a magnetic field of $\sim$ $2.4\times10^{12}$ Gauss in the emitting region. Under a magnetic field of $\sim$ $10^{12}$ Gauss, the ratio of cyclotron absorption coefficient between the fundamental absorption line and first harmonic is $\sim 10^{-1}-10^{-2}$ (You et al. 1997), which means that the chance of producing the fundamental absorption line is 10--100 times larger than producing the first harmonic. In case of a relatively low accretion rate, there may not be sufficient electrons near the surface of the neutron star to produce the first harmonics in the spectrum. If the accretion rate is high and more electrons are available, the first harmonic may appear in  the spectrum, and even the second or third harmonics might. \igr only shows fundamental absorption line and there is no sight of any harmonics, which hints to a  relatively low state of accretion. In addition to its single peak pulse profile, this is another indication that \igr is not in a very high accretion state during this outburst.

\section*{Acknowledgments}

We acknowledge support from the National Natural Science Foundation of
China via NSFC-10325313, 10521001, 10733010, 11073021,
11133002, 10821061, 10725313 ,the CAS key Project KJCX2-
YW-T03, and 973 program 2009CB824800.
Y.P.C. thanks the Natural Science Foundation of China for support via NSFC-11103020 and 11133002.
DFT acknowledges supported by the grants AYA2009-07391 and SGR2009-811, as well as the Formosa programme TW2010005 and iLINK programme 2011-0303. Shu Zhang would like to thank \emph{INTEGRAL} and \emph{Swift} for approving \emph{INTEGRAL} Data Right proposal No. 0820024 and \emph{Swift} ToO (Target of Opportunity) proposal (ID No. 35700), and for subsequently carrying out the observations to support this research.

\end{document}